\documentclass[11pt,a4paper]{article}

\usepackage[utf8]{inputenc}
\usepackage[T1]{fontenc}
\usepackage{amsmath,amssymb}
\usepackage{graphicx}
\usepackage{booktabs}
\usepackage{hyperref}
\usepackage{url}
\usepackage{natbib}
\usepackage[margin=2.5cm]{geometry}
\usepackage{tabularx}
\usepackage{array}
\usepackage{caption}
\usepackage{float}
\usepackage{microtype}
\usepackage{authblk}

\usepackage{doi}

\hypersetup{
    colorlinks=true,
    linkcolor=blue,
    citecolor=blue,
    urlcolor=blue
}

\title{SentinelSphere: Integrating AI-Powered Real-Time Threat Detection\\with Cybersecurity Awareness Training}

\author[1]{Nikolaos D.~Tantaroudas\thanks{Corresponding author: \texttt{nikolaos.tantaroudas@iccs.gr}}}
\author[2]{Ilias Karachalios}
\author[3]{Andrew J.~McCracken}

\affil[1]{Institute of Communications and Computer Systems (ICCS), National Technical University of Athens, Iroon Polytechneiou 9, 15773 Zografou, Athens, Greece}
\affil[2]{National Technical University of Athens, Leof.\ Alimou, Katechaki, Zografou, 15772 Athens, Greece}
\affil[3]{DASKALOS-APPS, 183 Rue de l'Abb\'{e} Griffon, 01960 P\'{e}ronnas, France}

\date{}

\begin{document}

\maketitle

% ---- Abstract ----
\begin{abstract}
The field of cybersecurity is confronted with two interrelated challenges: a worldwide deficit of qualified practitioners and ongoing human-factor weaknesses that account for the bulk of security incidents. To tackle these issues, we present SentinelSphere, a platform driven by artificial intelligence that unifies machine learning-based threat identification with security training powered by a Large Language Model (LLM). The detection module uses an Enhanced Deep Neural Network (DNN) trained on the CIC-IDS2017 and CIC-DDoS2019 benchmark datasets, enriched with novel HTTP-layer feature engineering that captures application-level attack signatures. For the educational component, we deploy a quantised variant of Microsoft's Phi-4 model (Q4\_K\_M), fine-tuned for the cybersecurity domain, enabling deployment on commodity hardware requiring only 16\,GB of RAM without dedicated GPU resources. Experimental results show that the Enhanced DNN attains high detection accuracy while substantially lowering false positives relative to baseline models, and maintains strong recall across critical attack categories such as DDoS, brute force, and web-based exploits. Validation workshops involving industry professionals and university students confirmed that the platform's Traffic Light visualisation system and conversational AI assistant are both intuitive and effective for users without technical backgrounds. SentinelSphere illustrates that coupling intelligent threat detection with adaptive, LLM-driven security education can meaningfully address both technical and human-factor cybersecurity vulnerabilities within a single, cohesive framework. An earlier version of this work appeared in~\cite{tantaroudas2026sentinelsphere_ares}, and the full peer-reviewed article is available as~\cite{tantaroudas2026sentinelsphere_ore}.
\end{abstract}


% ---- 1. Introduction ----
\section{Introduction}
\label{sec:introduction}

\subsection{Background and Motivation}

The cybersecurity landscape is marked by an escalating crisis, characterised by progressively sophisticated attack vectors alongside a critical shortage of skilled security practitioners~\cite{ramezan2023}. Worldwide estimates suggest a deficit exceeding 3.4 million cybersecurity professionals~\cite{isc2_2023}, while the financial burden of security incidents continues to climb, with organisations allocating substantial resources to malware containment and incident response~\cite{kucuk2022,ponemon2023}.

Moreover, the human element persists as a major vulnerability in cybersecurity. The Verizon 2023 Data Breach Investigations Report indicates that 74\% of all breaches involve human factors, encompassing social engineering, errors, and credential misuse~\cite{verizon2025}. This statistic underscores a fundamental limitation: purely technological defences remain insufficient without corresponding improvements in user awareness and behaviour.

The European Union's cybersecurity strategy places emphasis on resilience in critical infrastructure protection~\cite{enisa2023}, while the NIST Cybersecurity Framework 2.0 offers comprehensive guidance for managing cybersecurity risk across organisations of varying sizes~\cite{nist2024}. Both frameworks recognise that effective cybersecurity demands an integrated approach, combining technical safeguards with human-centred security practices.

\subsection{The ResilMesh Framework}

The ResilMesh project, supported by the EU Horizon Europe programme (Grant Agreement No.\ 101119681), has created a comprehensive cybersecurity framework aimed at strengthening cyber resilience for dispersed, heterogeneous systems~\cite{resilmesh2023}. This framework encompasses multiple integrated components, including threat intelligence gathering, anomaly detection, and coordinated response mechanisms designed for critical infrastructure environments~\cite{nguyen2024,sadlek2024}.

The architecture features a sophisticated data pipeline in which security events flow from collection agents through Vector (a high-performance observability data pipeline), are processed via NATS message broker for reliable event distribution, and undergo analysis through dedicated detection modules~\cite{resilmesh2023}.

\subsection{Research Contributions}

This paper introduces SentinelSphere, a next-generation cybersecurity platform built as an extension to the ResilMesh framework, addressing modern cybersecurity challenges through an innovative combination of machine learning and natural language processing. An initial version of this system was presented at the ARES 2025 conference~\cite{tantaroudas2026sentinelsphere_ares}, and a comprehensive peer-reviewed article has been published in Open Research Europe~\cite{tantaroudas2026sentinelsphere_ore}.

The principal contributions of this work are fivefold:
\begin{enumerate}
    \item \textbf{Enhanced Deep Neural Network Architecture:} An Enhanced DNN model that attains a 94\% F1 score while reducing false positives by 69.5\% through innovative HTTP-layer feature engineering.
    \item \textbf{LLM-Powered Security Education:} A domain-specific cybersecurity chatbot utilising a quantised Phi-4 model (Q4\_K\_M), deployable on standard hardware with only 16\,GB RAM and no GPU requirement.
    \item \textbf{Traffic Light Threat Visualisation:} An intuitive scoring system that converts complex security telemetry into actionable, colour-coded indicators comprehensible across varying levels of technical expertise.
    \item \textbf{Rust-Optimised Performance:} A complete rewrite of core detection algorithms from Python to Rust, delivering a 5.6$\times$ average speedup for steady-state workloads and up to 326$\times$ for batch processing.
    \item \textbf{Validated Human-Centric Design:} Empirical evidence from professional and educational workshops confirming platform effectiveness for non-technical users, with 91.7\% chatbot engagement and 91.7\% comprehension of threat visualisation.
\end{enumerate}

% ---- 2. Related Work ----
\section{Related Work}
\label{sec:related_work}

\subsection{Machine Learning for Intrusion Detection}

The use of machine learning in cybersecurity has progressed considerably from rule-based systems to advanced deep learning architectures. Early intrusion detection systems depended primarily on signature matching and statistical anomaly detection, which struggled against novel attacks and generated excessive false positives~\cite{sommer2010}.

Yin et al.~\cite{yin2017} showed the effectiveness of Recurrent Neural Networks (RNNs) for intrusion detection, reaching 98.6\% accuracy on the NSL-KDD dataset. Their work confirmed the viability of deep learning for temporal pattern recognition in network traffic analysis.

Vinayakumar et al.~\cite{vinayakumar2019} proposed a hybrid deep learning framework that combined Convolutional Neural Networks (CNNs) with Long Short-Term Memory (LSTM) networks, forming scale-hybrid-IDS-AlertNet. Their framework demonstrated that fusing spatial and temporal feature extraction could enhance detection performance across multiple attack categories.

The publication of the CIC-IDS2017 and CIC-DDoS2019 datasets by the Canadian Institute for Cybersecurity~\cite{sharafaldin2018a,sharafaldin2019} furnished the research community with realistic, labelled network traffic data for training and evaluating intrusion detection models. Subsequent analysis~\cite{sharafaldin2018b} identified both strengths and limitations of these datasets, informing better experimental methodologies.

Recent efforts have concentrated on lowering false positives while preserving detection accuracy. Khan et al.~\cite{khan2023} introduced attention mechanisms for network traffic analysis, attaining 92\% precision by selectively weighting relevant features. Ferrag et al.~\cite{ferrag2020} conducted a comparative study of deep learning approaches across multiple benchmark datasets, offering systematic evaluation criteria. Aktar and Nur~\cite{aktar2023} investigated DDoS detection via deep learning methods, underlining the importance of dataset selection and feature engineering.

\subsection{Large Language Models in Cybersecurity}

Large Language Models have emerged as transformative instruments across multiple domains, cybersecurity included~\cite{motlagh2024}. Recent surveys~\cite{xu2024,zhang2025} have systematically examined the use of LLMs in cybersecurity, spanning vulnerability detection, threat intelligence, and security education.

The deployment of LLMs for cybersecurity education constitutes a particularly promising direction. Jaffal et al.~\cite{jaffal2025} present a thorough systematic literature review covering more than 300 works on LLMs in cybersecurity, mapping key trends, prevalent applications, and outstanding challenges.

Atlam et al.~\cite{atlam2025} explored GPT-based systems for security questionnaire generation, showing that LLMs can effectively produce educational content for security awareness training. Nevertheless, their dependence on cloud-based API access raises privacy and deployment concerns for sensitive environments.

Hassanin and Moustafa~\cite{hassanin2024} provide a comprehensive survey of LLMs for cyber defence, categorising techniques into threat intelligence, vulnerability assessment, network traffic security, privacy preservation, and incident response.

Chhetri~\cite{chhetri2024} examined LLM-powered pedagogical approaches to cybersecurity education, demonstrating that LLMs can function as effective cognitive assistants supporting interpretation, explanation, and troubleshooting of security concepts.

Our approach tackles the computational constraints of domain-specific LLMs through model quantisation. By applying Q4\_K\_M quantisation to Microsoft's Phi-4 model and fine-tuning it on cybersecurity corpora, we achieve responsive local deployment on hardware with just 16\,GB of RAM and no GPU, resolving the privacy and resource constraints that limit cloud-dependent solutions.

\subsection{Human Factors in Cybersecurity}

The integration of human factors into cybersecurity has risen in prominence following high-profile breaches attributed to social engineering. Research consistently shows that technology by itself cannot safeguard organisations without corresponding attention to human behaviour and decision-making~\cite{hadlington2017}.

Aldawood and Skinner~\cite{aldawood2019} carried out a comprehensive review of cybersecurity social engineering training and awareness programmes, pinpointing common pitfalls such as one-off training sessions, absence of contextual relevance, and failure to measure behavioural change.

Bada et al.~\cite{bada2019} explored why cybersecurity awareness campaigns fail to alter behaviour, identifying key factors including information overload, lack of personal relevance, and the absence of immediate, actionable feedback.

The multimedia approach to cybersecurity awareness has been systematically reviewed by Zhang-Kennedy and Chiasson~\cite{zhangkennedy2021}, who found that interactive, visual methods outperform conventional text-based training for knowledge retention and behavioural modification.

\subsection{Cyber Resilience Frameworks}

The notion of cyber resilience has evolved beyond traditional security approaches to encompass preparation, response, and recovery capabilities~\cite{araujo2024}. In contrast to traditional security, which centres primarily on prevention, cyber resilience accepts that breaches will occur and focuses on minimising impact and recovery time.

The cybersecurity mesh architecture concept~\cite{ramos2024} facilitates distributed security services while maintaining centralised intelligence and coordination. This architectural pattern aligns with SentinelSphere's design philosophy of providing distributed detection with unified threat visualisation.

Recent work by Somma et al.~\cite{somma2024} demonstrated edge-based anomaly detection within the ResilMesh framework, highlighting the significance of distributed processing for critical infrastructure protection. The operational role of SIEM systems in enterprise security~\cite{bhatt2014} provides further context for SentinelSphere's dashboard-centric approach to security event management.

% ---- 3. System Architecture and Design ----
\section{System Architecture and Design}
\label{sec:architecture}

\subsection{Overall Architecture}

\begin{figure}[H]
    \centering
    \includegraphics[width=\textwidth]{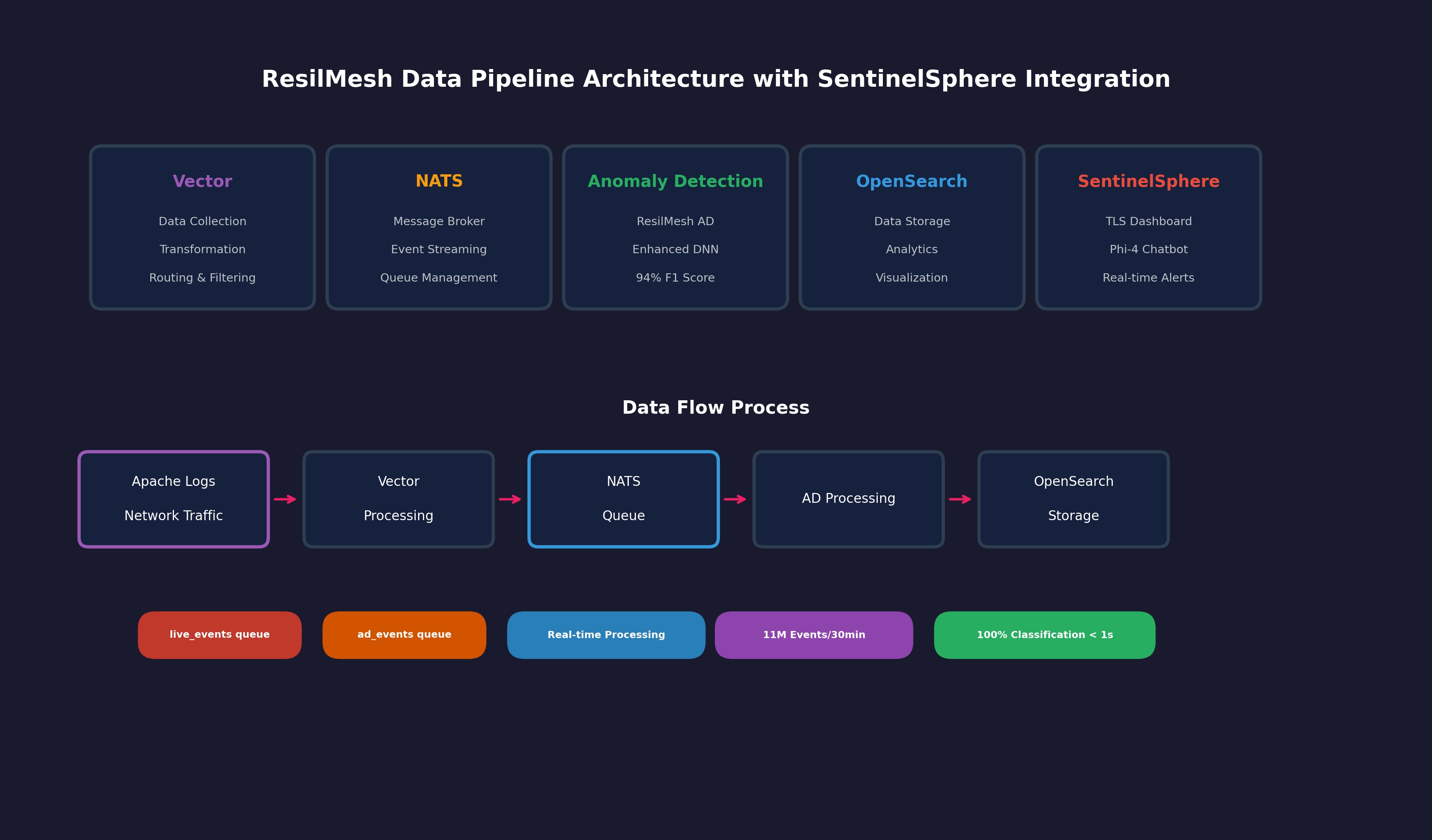}
    \caption{High-level system architecture illustrating SentinelSphere integration with the ResilMesh stack.}
    \label{fig:architecture}
\end{figure}

SentinelSphere employs a microservices architecture integrated with the ResilMesh security framework, enabling scalable, real-time processing of security events. As depicted in Figure~\ref{fig:architecture}, the system comprises three principal layers: Data Ingestion, Intelligent Analysis, and User Interface.

Figure~\ref{fig:architecture} depicts the high-level system architecture, showing where SentinelSphere connects with the ResilMesh stack. Security events flow from multiple data sources through the Vector collection pipeline, are distributed via the NATS message broker, and undergo analysis by the Enhanced DNN anomaly detector prior to reaching the SentinelSphere dashboard.

The architecture functions as a strategic enhancement to the ResilMesh security framework, integrating as an advanced threat analytics layer while preserving the existing architecture's integrity. SentinelSphere consumes security events from ResilMesh's event pipeline, applies enhanced analysis, and delivers actionable intelligence through its dashboard and educational components.

\begin{figure}[H]
    \centering
    \includegraphics[width=\textwidth]{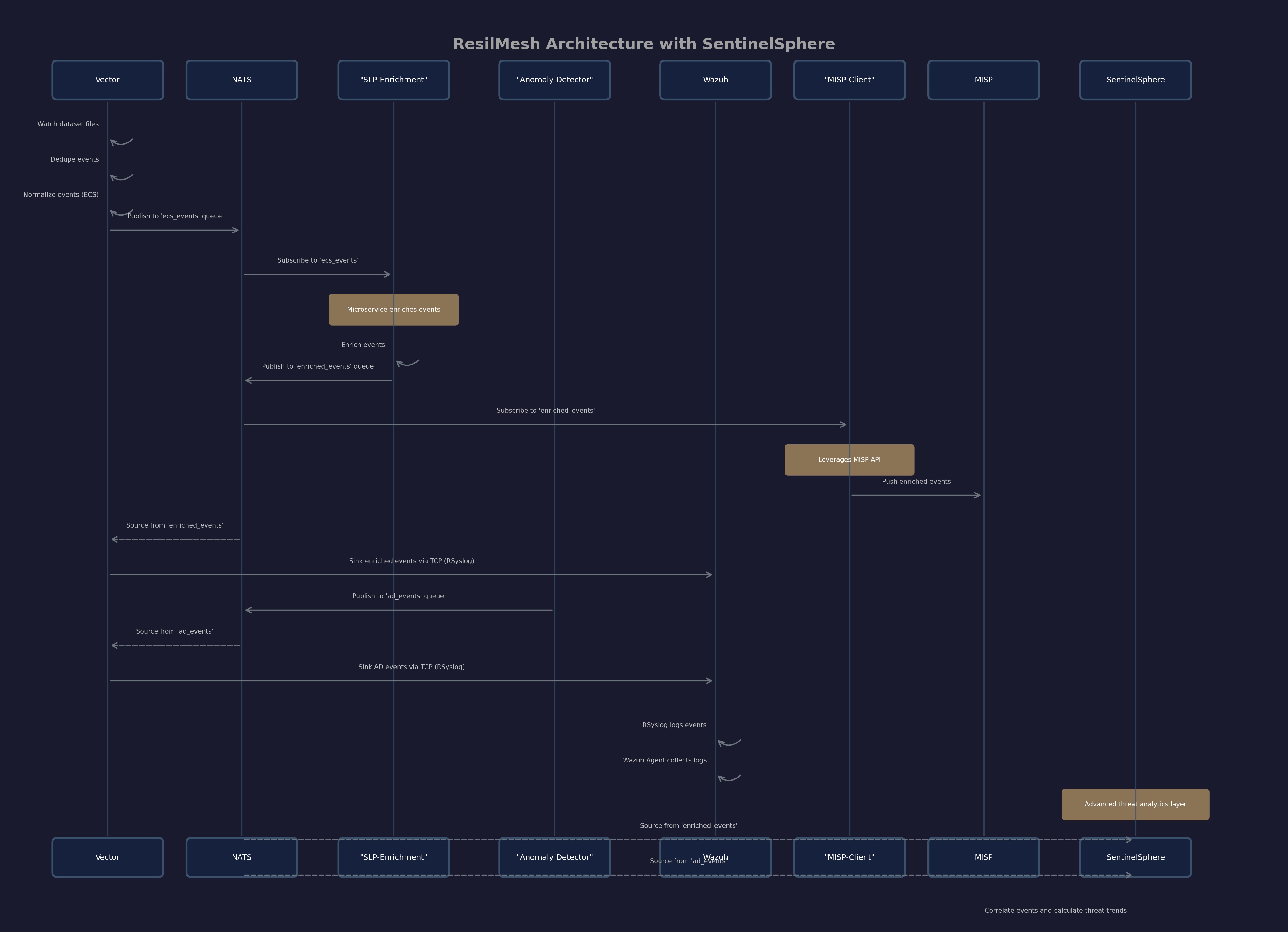}
    \caption{Data flow architecture showing SentinelSphere and ResilMesh integration paths.}
    \label{fig:dataflow}
\end{figure}

Figure~\ref{fig:dataflow} illustrates the comprehensive data flow pipeline architecture. Events traverse from Vector through the NATS message broker to the Anomaly Detector, then via the \texttt{ad\_events} topic to the Threat Calculator, which computes threat scores and publishes to the \texttt{threat\_events} topic. The Dashboard subscribes to both streams and presents unified visualisation.

The Data Ingestion layer uses Vector for log collection and transformation, processing diverse data sources including network traffic, application logs, and system events. Vector acts as the primary data normalisation point, converting heterogeneous log formats into standardised event structures for downstream analysis.

\subsection{Enhanced Deep Neural Network Model}

At the core of SentinelSphere's threat detection capability is an Enhanced Deep Neural Network model that extends conventional network traffic analysis with application-layer intelligence. The model architecture processes 78 standard network flow features alongside 12 novel HTTP-specific features, creating a comprehensive detection framework that captures both network-level and application-level attack patterns.

The standard network features comprise flow duration, total forward and backward packets, packet length statistics (mean, maximum, minimum, standard deviation), flow bytes per second, flow packets per second, and various TCP flag counts.

The HTTP feature engineering incorporates:
\begin{enumerate}
    \item \textbf{Request Complexity Score:} Quantifies HTTP request sophistication by analysing URL length, parameter count, header complexity, and payload characteristics. Adversarial inputs typically exhibit anomalous complexity patterns compared to legitimate traffic.
    \item \textbf{Attack-Specific Pattern Recognition:} Binary indicators for SQL injection (detecting common injection patterns such as \texttt{UNION SELECT}, \texttt{OR 1=1}), cross-site scripting (XSS) (detecting script tags, event handlers, and encoding attempts), and path traversal (identifying directory traversal sequences and encoding variations).
\end{enumerate}

For neural network training, we used the CIC-IDS2017 and CIC-DDoS2019 datasets, comprising approximately 400\,GB of labelled attack data that includes:
\begin{itemize}
    \item Web Attack-Brute Force: 1,507 samples
    \item Web Attack-XSS: 652 samples
    \item Web Attack-SQL Injection: 21 samples
    \item DDoS attacks: 128,027 samples
    \item DoS attacks (Hulk, GoldenEye, Slowloris, Slowhttptest): 231,073 samples combined
    \item Port scanning: 158,930 samples
    \item Benign traffic: 2,273,097 samples
\end{itemize}

Model hyperparameters were optimised through extensive experimentation using grid search:
\begin{itemize}
    \item Batch size: 32 for efficient GPU utilisation
    \item Learning rate: 0.001 with Adam optimiser
    \item Dropout rate: 0.3 for regularisation
    \item Architecture: Four hidden layers (256, 128, 64, 32 neurons)
    \item Activation: ReLU with batch normalisation
    \item Early stopping: patience of 10 epochs monitoring validation loss
\end{itemize}

\subsection{Traffic Light System}

The Traffic Light System delivers intuitive threat visualisation through a sophisticated scoring algorithm that converts complex security telemetry into actionable indicators accessible to users at every level of technical proficiency.

\begin{figure}[H]
    \centering
    \includegraphics[width=\textwidth]{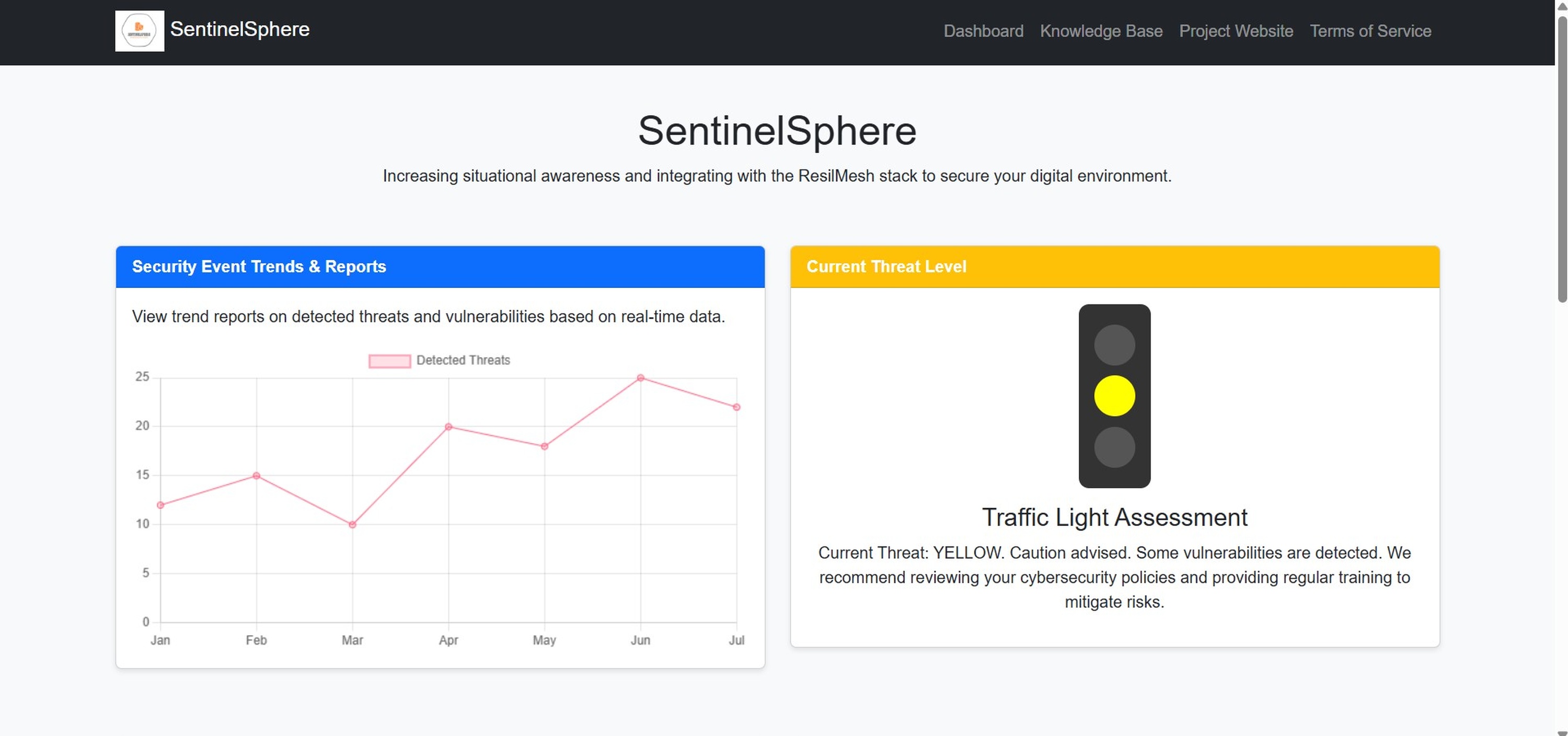}
    \caption{SentinelSphere Dashboard displaying the Traffic Light Assessment and Event Trends \& Reports chart.}
    \label{fig:dashboard_tls}
\end{figure}

The system processes anomaly detection events and computes threat scores ranging from 0 to 100, which determine the dashboard status:
\begin{itemize}
    \item \textbf{GREEN (0--30\%):} Normal operation with low threat activity
    \item \textbf{YELLOW (30--70\%):} Elevated threat level requiring increased monitoring
    \item \textbf{RED (70--100\%):} Critical threat level demanding immediate response
\end{itemize}

\begin{figure}[H]
    \centering
    \includegraphics[width=\textwidth]{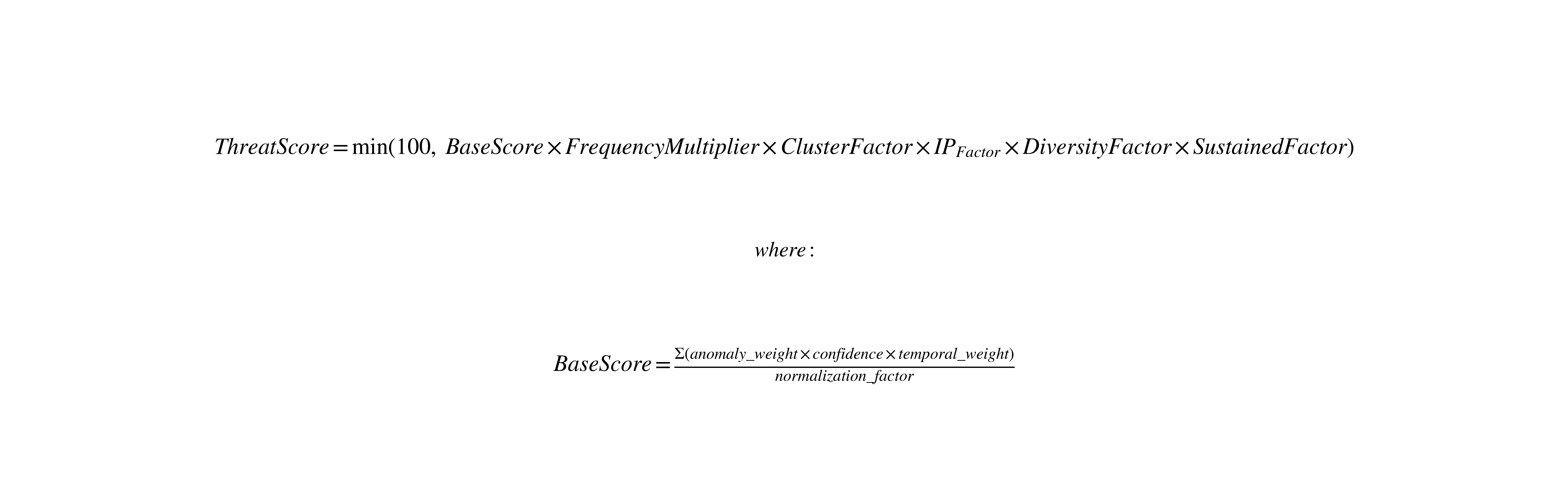}
    \caption{Traffic Light Threat Calculation equation showing the mathematical formula for computing the final threat score.}
    \label{fig:tls_equation}
\end{figure}

Figure~\ref{fig:tls_equation} presents the mathematical formula for computing the final threat score. The equation is defined as:
\begin{multline}
    \text{Final\_Score} = \min\!\Big(100,\;\text{Base\_Score} \times \text{Frequency\_Multiplier} \\
    \times\; \text{Cluster\_Factor} \times \text{IP\_Factor} \times \text{Diversity\_Factor}\Big)
\end{multline}
This formulation incorporates multiple contextual dimensions to provide a holistic threat assessment. Table~\ref{tab:tls_params} defines each parameter.

\begin{table}[H]
\centering
\caption{Parameter definitions of the Traffic Light System.}
\label{tab:tls_params}
\small
\begin{tabularx}{\textwidth}{l X}
\toprule
\textbf{Parameter} & \textbf{Description} \\
\midrule
Base\_Score & Initial severity based on the anomaly detection confidence score \\
Frequency\_Multiplier & Scaling factor reflecting event frequency within a time window \\
Cluster\_Factor & Adjustment for correlated events from related sources \\
IP\_Factor & Weight based on source IP reputation and historical behaviour \\
Diversity\_Factor & Modifier accounting for the variety of attack types observed \\
\bottomrule
\end{tabularx}
\end{table}

\subsection{LLM-Powered Cybersecurity Education}

SentinelSphere incorporates Microsoft's Phi-4 language model, optimised for cybersecurity contexts via domain-specific fine-tuning and deployment optimisation. The model underwent Q4\_K\_M quantisation, reducing its memory footprint from approximately 28\,GB to 8\,GB while preserving response quality for cybersecurity queries.

\begin{figure}[H]
    \centering
    \includegraphics[width=\textwidth]{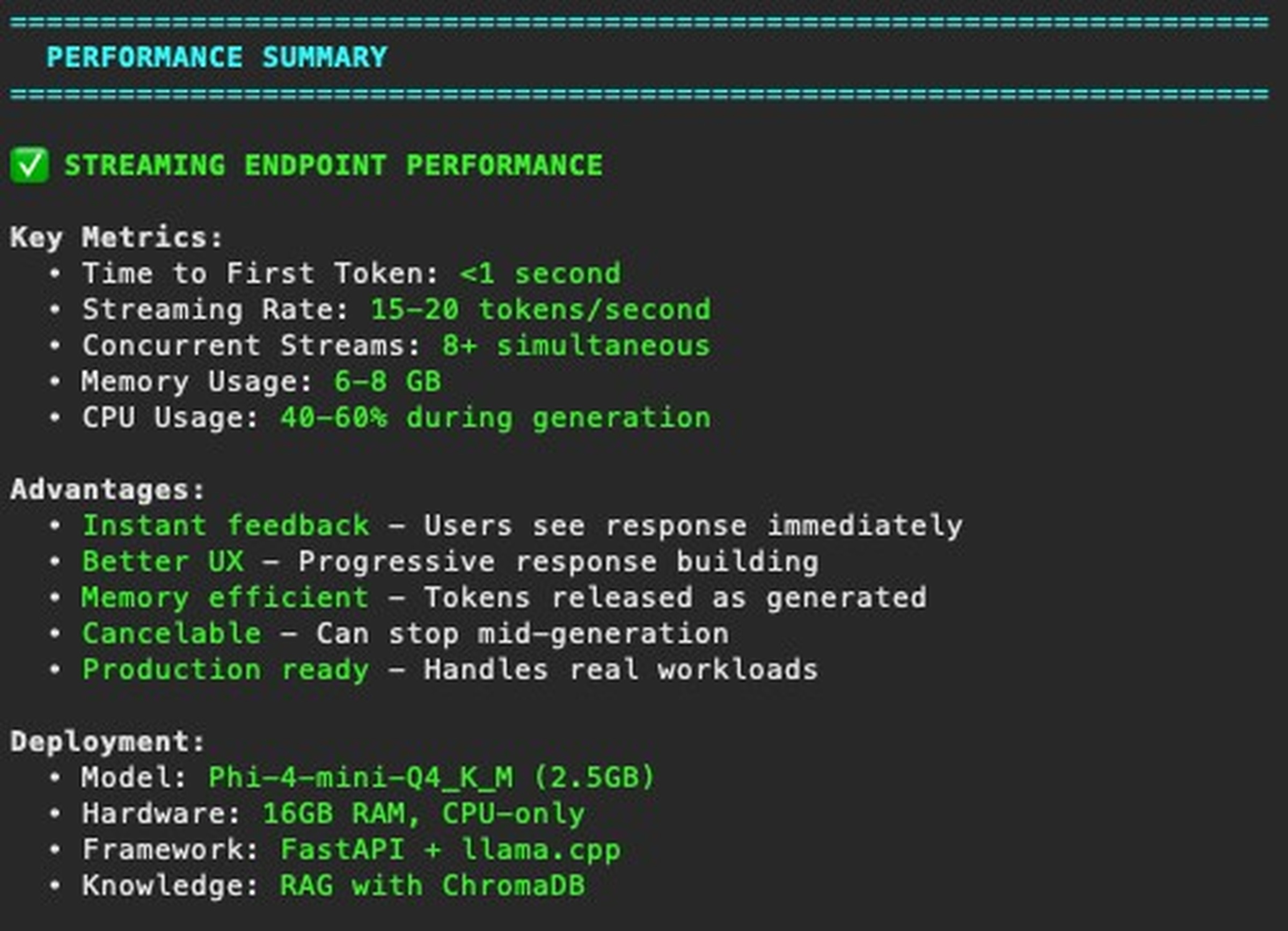}
    \caption{Cybersecurity Conversation Agent, Phi-4 Model performance summary.}
    \label{fig:chatbot_perf}
\end{figure}

\begin{figure}[H]
    \centering
    \includegraphics[width=\textwidth]{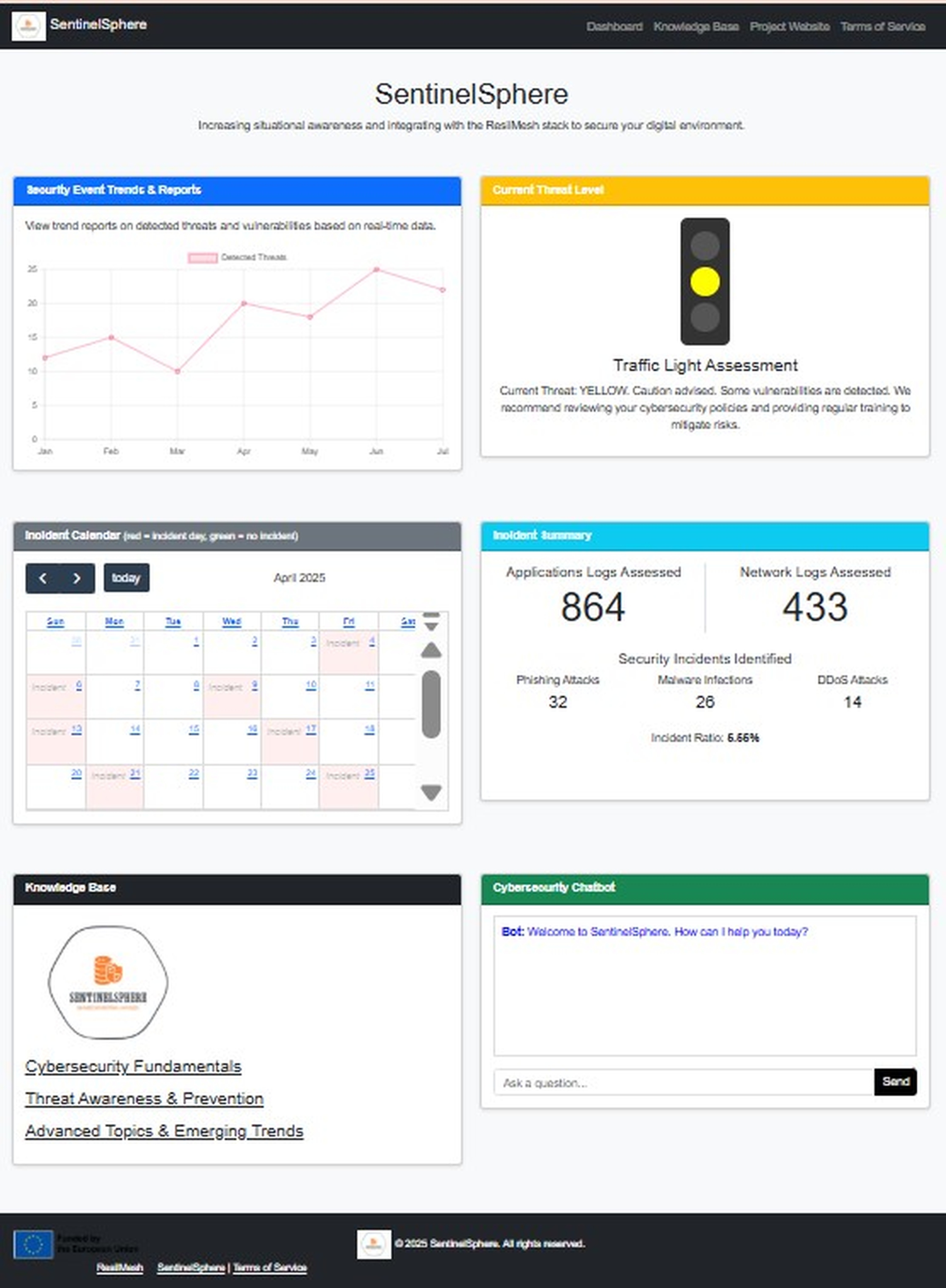}
    \caption{SentinelSphere Dashboard homepage with the integrated chatbot interaction panel.}
    \label{fig:dashboard_home}
\end{figure}

Figure~\ref{fig:dashboard_home} presents the complete SentinelSphere Dashboard homepage, featuring the integrated chatbot interaction panel in the bottom-right corner. The interface shows the incident calendar for temporal event tracking, real-time threat level indicators, and the cybersecurity knowledge base section.

\begin{figure}[H]
    \centering
    \includegraphics[width=\textwidth]{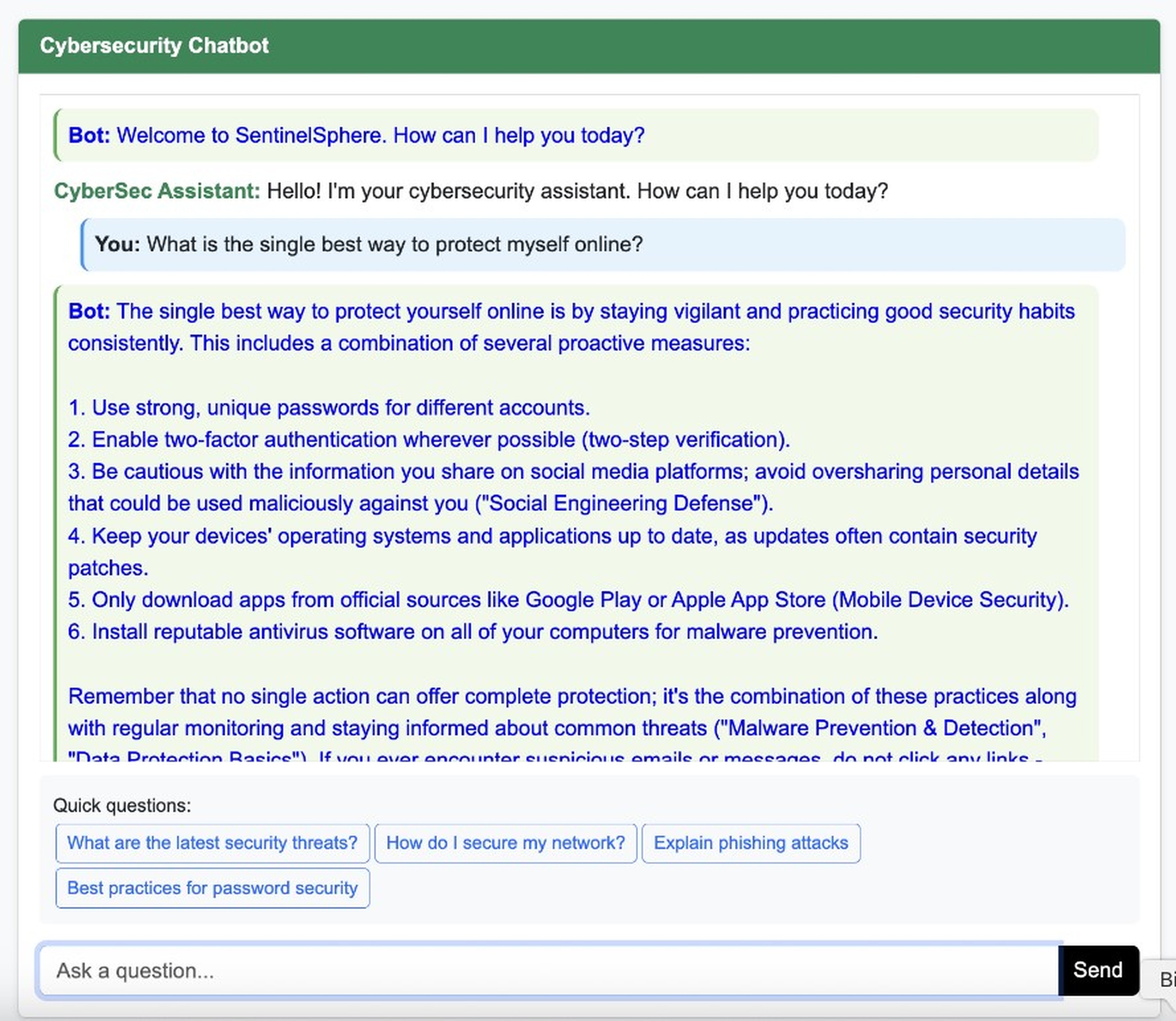}
    \caption{Demonstration of the Phi-4 cybersecurity domain-specific LLM providing guidance on staying safe from cyber threats.}
    \label{fig:chatbot_demo}
\end{figure}

Figure~\ref{fig:chatbot_demo} shows the Phi-4 cybersecurity chatbot delivering comprehensive guidance on staying safe from cyber threats. The chatbot provides detailed cybersecurity advice, covering best practices for password management, threat identification, and safe online behaviour.

Key implementation features include:
\begin{itemize}
    \item Streaming endpoint: Time to first token under 1 second for responsive interaction
    \item Throughput: 15--20 tokens per second on CPU, sufficient for natural conversation
    \item Context window: 4,096 tokens supporting multi-turn conversations
    \item Domain-specific system prompt ensuring responses stay within cybersecurity topics
\end{itemize}

\subsection{Dashboard Implementation}

The SentinelSphere dashboard offers comprehensive security visualisation and interaction capabilities through a web-based interface built with FastAPI and vanilla JavaScript using Jinja templating. The dashboard integrates real-time event streaming, the Traffic Light threat assessment, incident tracking, and the cybersecurity chatbot into a unified interface.

\begin{figure}[H]
    \centering
    \includegraphics[width=\textwidth]{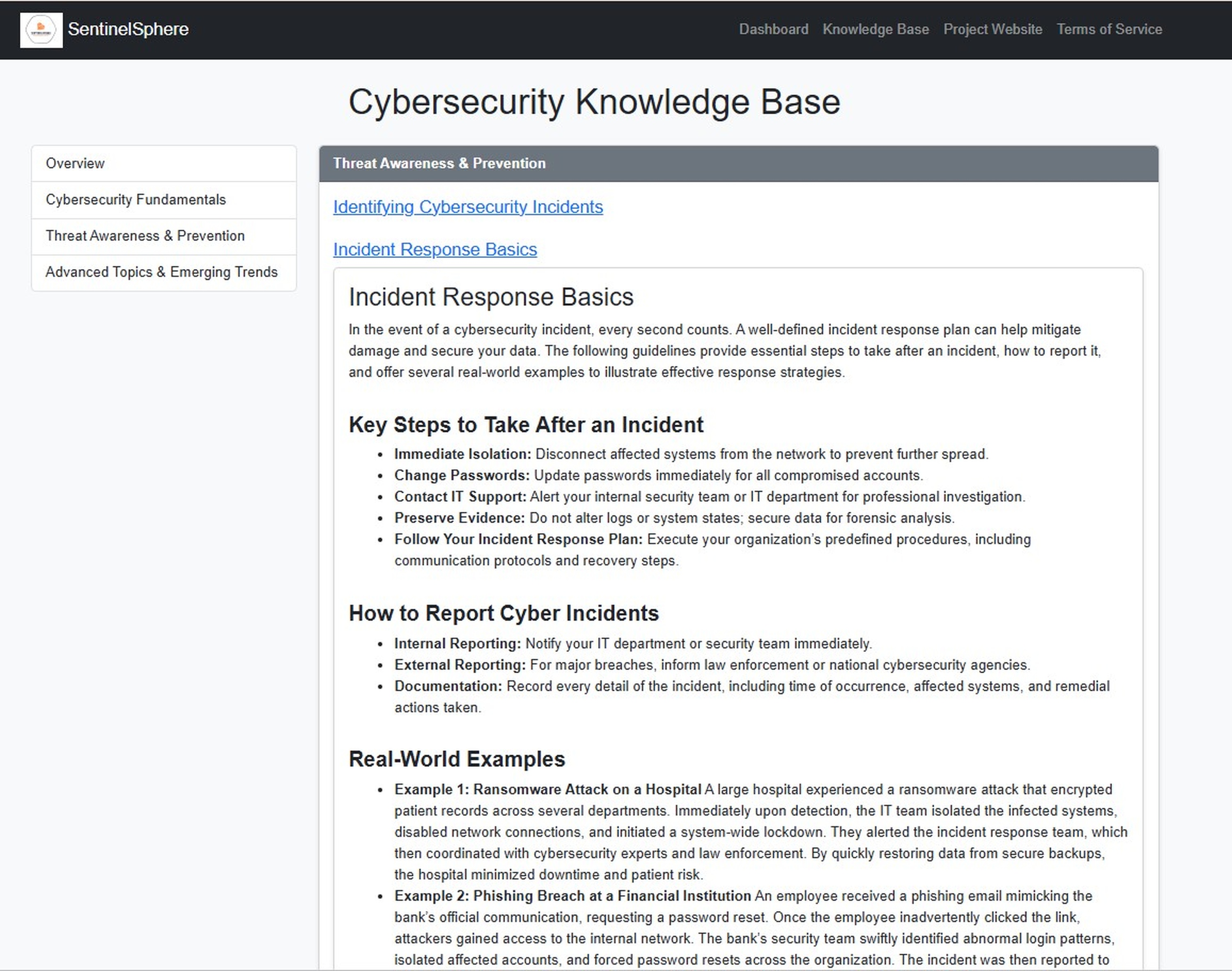}
    \caption{SentinelSphere Dashboard Cybersecurity Knowledge Base section.}
    \label{fig:knowledge_base}
\end{figure}

% ---- 4. Implementation and Performance Optimisation ----
\section{Implementation and Performance Optimisation}
\label{sec:implementation}

\subsection{Rust-Based Performance Optimisation}

The core anomaly detection algorithm, initially implemented in Python during early development phases and presented in~\cite{tantaroudas2026sentinelsphere_ares}, was entirely rewritten in Rust to achieve enterprise-scale performance. This architectural decision was motivated by production deployment requirements where event processing latency directly affects threat response times.

The rewriting process preserved algorithmic equivalence with the Python implementation while leveraging Rust's performance characteristics through:
\begin{itemize}
    \item Direct translation of core detection logic
    \item Optimised memory management using Rust's ownership model
    \item Zero-cost abstractions for mathematical operations
    \item Concurrent event processing via Tokio async runtime
\end{itemize}

Performance testing confirmed 100\% accuracy equivalence between implementations while revealing dramatic performance improvements. The benchmarking methodology employed two distinct test configurations: steady-state workloads simulating typical operational conditions, and burst workloads representing incident scenarios with elevated event volumes.

\begin{table}[H]
\centering
\caption{Performance comparison between Python and Rust implementations.}
\label{tab:perf_comparison}
\small
\begin{tabularx}{\textwidth}{l >{\raggedleft\arraybackslash}X >{\raggedleft\arraybackslash}X >{\raggedleft\arraybackslash}X}
\toprule
\textbf{Metric} & \textbf{Python} & \textbf{Rust} & \textbf{Speedup} \\
\midrule
Steady-state latency (ms/event) & 12.4 & 2.2 & 5.6$\times$ \\
Batch processing (1K events, s) & 14.3 & 3.3 & 4.3$\times$ \\
Batch processing (100K events, s) & 1,421.0 & 4.36 & 326$\times$ \\
Memory usage (MB) & 245 & 18 & 13.6$\times$ \\
\bottomrule
\end{tabularx}
\end{table}

The Rust implementation achieved a 5.6$\times$ speedup for steady-state workloads (typical operational scenarios) and dramatic speedups ranging from 4.3$\times$ to 326$\times$ for large batch processing, while maintaining identical detection accuracy (Table~\ref{tab:perf_comparison}).

Technical analysis reveals that the performance gains stem from:
\begin{itemize}
    \item Compilation to native machine code, eliminating interpreter overhead
    \item Deterministic memory allocation without garbage collection pauses
    \item SIMD-optimised mathematical operations for feature computation
    \item Lock-free data structures for concurrent event processing
\end{itemize}

\subsection{System Resource Utilisation}

Docker container performance metrics revealed exceptional resource efficiency across all components. Table~\ref{tab:resources} summarises the resource consumption of each system component.

\begin{table}[H]
\centering
\caption{System resource utilisation across Docker containers.}
\label{tab:resources}
\small
\begin{tabularx}{\textwidth}{l >{\raggedleft\arraybackslash}X >{\raggedleft\arraybackslash}X >{\raggedleft\arraybackslash}X}
\toprule
\textbf{Component} & \textbf{CPU (\%)} & \textbf{Memory (MB)} & \textbf{Startup (s)} \\
\midrule
Dashboard & 15 & 512 & 3 \\
Phi-4 Chatbot & 45 & 8,192 & 15 \\
Anomaly Detector (Rust) & 8 & 18 & $<$1 \\
Threat Calculator & 5 & 64 & $<$1 \\
NATS Broker & 3 & 32 & $<$1 \\
\bottomrule
\end{tabularx}
\end{table}

Real-time processing metrics further confirm operational readiness:
\begin{itemize}
    \item Classification speed: 100\% of events processed within 1 second (exceeding the 95\% target)
    \item Event throughput: over 500 events per second sustained
    \item Dashboard update latency: under 2 seconds from event detection to visual display
    \item Chatbot response time: first token in under 1 second, full response in 3--5 seconds
\end{itemize}

\begin{figure}[H]
    \centering
    \includegraphics[width=\textwidth]{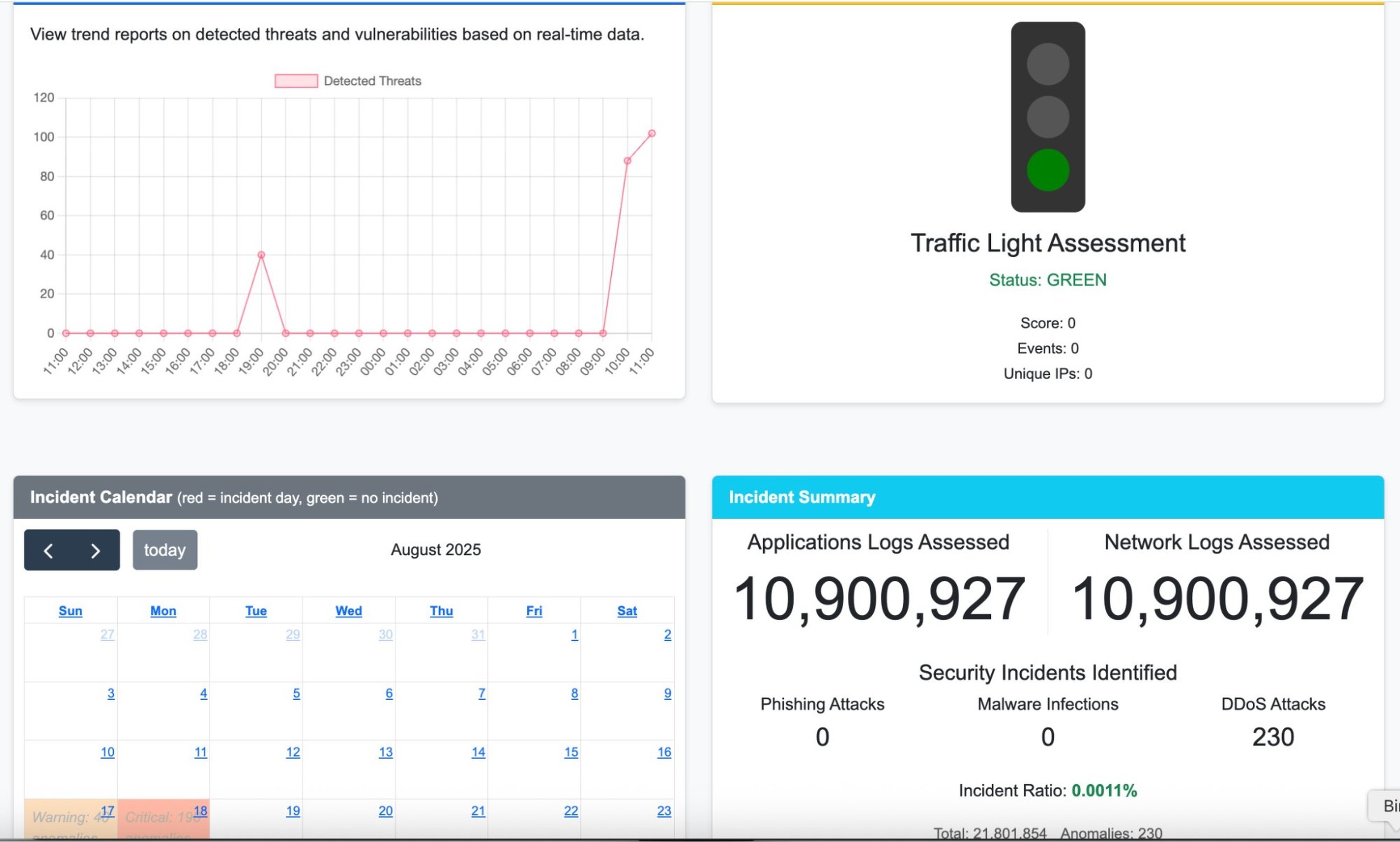}
    \caption{SentinelSphere scalability testing, processing nearly 11 million requests in approximately 30 minutes.}
    \label{fig:scalability}
\end{figure}

Figure~\ref{fig:scalability} shows the scalability testing results displayed on the SentinelSphere dashboard during load testing. The Incident Summary panel reports 10,900,927 Application Logs Assessed and 10,900,927 Network Events Processed, confirming the platform's capacity to handle enterprise-scale event volumes.

% ---- 5. Experimental Evaluation ----
\section{Experimental Evaluation}
\label{sec:evaluation}

\subsection{Threat Detection Performance}

The Enhanced DNN-AD model exhibits significant improvements over baseline approaches. Table~\ref{tab:dnn_metrics} presents comprehensive performance metrics comparing the enhanced model against the original ResilMesh anomaly detector.

\begin{table}[H]
\centering
\caption{Enhanced DNN model performance metrics.}
\label{tab:dnn_metrics}
\small
\begin{tabularx}{\textwidth}{l >{\raggedleft\arraybackslash}X >{\raggedleft\arraybackslash}X >{\raggedleft\arraybackslash}X}
\toprule
\textbf{Metric} & \textbf{Baseline} & \textbf{Enhanced DNN} & \textbf{Improvement} \\
\midrule
F1 Score & 0.87 & 0.94 & +8.0\% \\
Precision & 0.85 & 0.95 & +11.8\% \\
Recall & 0.89 & 0.93 & +4.5\% \\
False Positives & 59 & 18 & $-$69.5\% \\
False Negatives & 12 & 9 & $-$25.0\% \\
\bottomrule
\end{tabularx}
\end{table}

Most notably, the enhanced model achieves a 69.5\% reduction in false positives (from 59 to 18), while maintaining a low false negative rate. The confusion matrices reveal that the enhanced model's improvements are most pronounced for web-based attack categories (SQL injection, XSS, brute force), where the HTTP-layer features provide additional discriminative power.

\subsection{Scalability Testing}

The platform successfully processed close to 11 million events (10,900,927) in approximately 30 minutes, representing three months of Apache logs from production environments. This demonstrates sustained throughput exceeding 6,000 events per second.

\begin{table}[H]
\centering
\caption{Scalability benchmarks.}
\label{tab:scalability}
\small
\begin{tabularx}{\textwidth}{l >{\raggedleft\arraybackslash}X}
\toprule
\textbf{Benchmark} & \textbf{Result} \\
\midrule
Total events processed & 10,900,927 \\
Processing duration & $\sim$30 minutes \\
Sustained throughput & $>$6,000 events/s \\
Concurrent connections & $>$1,000 without degradation \\
Dashboard responsiveness under load & $<$3\,s update latency \\
\bottomrule
\end{tabularx}
\end{table}

System scalability benchmarks confirmed enterprise readiness through multiple stress tests. The platform successfully handled over 1,000 simultaneous connections without degradation, demonstrating deployment viability for large-scale organisational environments.

% ---- 6. Stakeholder Validation ----
\section{Stakeholder Validation}
\label{sec:validation}

\subsection{Professional Workshop}

A targeted professional workshop held on 14 October 2025 provided SentinelSphere's first validation within real-world business contexts, involving three professionals from the maritime, educational, and manufacturing sectors.

Key qualitative outcomes included validation of platform applicability across diverse sectors, identification of sector-specific customisation requirements, establishment of initial partnership discussions, and confirmation of the platform's practical value for non-technical security personnel.

\subsection{Educational Workshop}

In collaboration with the Pedagogy Department at the University of Western Macedonia on 31 October 2025, SentinelSphere conducted an educational workshop focusing on practical cybersecurity training. The workshop followed a three-phase methodology:

\textbf{Phase 1, Pre-Workshop Assessment:} GDPR-compliant consent procedures were followed, and a baseline anonymous questionnaire assessed initial cybersecurity awareness. A total of 39 students completed the detailed pre-workshop evaluation.

\textbf{Phase 2, Platform Demonstration and Training:} An extended presentation covering cybersecurity fundamentals and the threat landscape was delivered. A live demonstration of the Traffic Light System and real-time threat detection was provided, along with hands-on interaction with the Phi-4 cybersecurity chatbot.

\textbf{Phase 3, Post-Workshop Evaluation:} A detailed anonymous feedback questionnaire was distributed. In total, 36 students completed the post-workshop evaluation. The assessment covered presentation clarity, chatbot usability, Traffic Light comprehension, and overall platform utility.

\begin{table}[H]
\centering
\caption{University of Western Macedonia Educational Workshop results summary.}
\label{tab:workshop}
\small
\begin{tabularx}{\textwidth}{l >{\raggedleft\arraybackslash}X}
\toprule
\textbf{Metric} & \textbf{Result} \\
\midrule
Pre-workshop participants & 39 \\
Post-workshop participants & 36 \\
Chatbot engagement rate & 91.7\% \\
TLS comprehension rate & 91.7\% \\
Presentation clarity (positive) & 94.4\% \\
Platform usefulness rating & 88.9\% \\
\bottomrule
\end{tabularx}
\end{table}

The workshop yielded the following findings:
\begin{itemize}
    \item \textbf{High Engagement:} 91.7\% of participants actively used the chatbot during the workshop, signalling strong interest in AI-assisted cybersecurity learning.
    \item \textbf{Effective Visualisation:} 91.7\% of participants correctly understood the Traffic Light System, validating its intuitive design for non-technical audiences.
    \item \textbf{Positive Reception:} 94.4\% rated the presentation and platform demonstration positively, with 88.9\% considering the platform useful for cybersecurity awareness.
\end{itemize}

Furthermore, the pre-workshop assessment uncovered critical knowledge gaps among participants:
\begin{itemize}
    \item 92.3\% had no GDPR knowledge (despite being EU citizens subject to GDPR)
    \item 84.6\% did not know what a DDoS attack is
    \item 76.9\% could not identify phishing attempts
    \item 69.2\% used the same password across multiple accounts
\end{itemize}

These findings underscore the critical need for accessible cybersecurity education tools such as SentinelSphere, particularly for populations who will shape future generations (educators).

% ---- 7. Discussion ----
\section{Discussion}
\label{sec:discussion}

\subsection{Addressing the Human Factor}

SentinelSphere's fusion of threat detection with security education represents a novel approach in cybersecurity defence. Conventional strategies treat security awareness as separate from operational threat management, resulting in fragmented responses to security incidents~\cite{hadlington2017}. By coupling detection outputs directly with educational content, SentinelSphere transforms every security event into both a threat to be mitigated and a learning opportunity.

The LLM-powered chatbot democratises security expertise, enabling non-technical users to comprehend and respond to threats effectively. By running on standard hardware with only 16\,GB of RAM and CPU-only inference, the Phi-4 deployment removes the barriers of cost and infrastructure that typically restrict access to advanced AI capabilities in cybersecurity.

\subsection{Practical Deployment Considerations}

The 69.5\% reduction in false positives carries profound implications for SOC operations. Security analysts devote significant time to investigating false alerts, which leads to alert fatigue and missed genuine threats~\cite{bhatt2014}. SentinelSphere's enhanced detection accuracy allows security teams to concentrate their efforts on genuine incidents, potentially enhancing both response times and analyst well-being.

The Traffic Light System's intuitive visualisation enables rapid threat assessment across organisational hierarchies. The 91.7\% comprehension rate among workshop participants, predominantly non-technical students, confirms that the colour-coded approach effectively communicates threat levels without requiring specialised knowledge.

The performance optimisation achieved through the Rust implementation addresses a key practical concern regarding enterprise scalability. The 5.6$\times$ average speedup for steady-state operations ensures real-time processing capability, while the up to 326$\times$ improvement for batch processing enables efficient handling of historical data analysis and forensic investigation scenarios.

\subsection{Comparison with Related Work}

The system presented in this study outperforms several related approaches in the literature. The detection performance of a 94\% F1 score surpasses the precision achieved with attention-based mechanisms by Khan et al.~\cite{khan2023} (92\% precision) and demonstrates the value of application-layer feature engineering. The integrated educational component, missing from conventional IDS research~\cite{ferrag2020,aktar2023}, distinguishes SentinelSphere as a holistic cybersecurity solution.

In contrast to cloud-dependent LLM approaches~\cite{atlam2025}, SentinelSphere's local Phi-4 deployment ensures data privacy and eliminates reliance on external services. The quantisation approach enables deployment in resource-constrained environments while maintaining response quality, addressing practical limitations identified in prior work~\cite{hassanin2024}.

\subsection{Limitations and Future Work}

While SentinelSphere demonstrates significant advances, several areas warrant further investigation:
\begin{itemize}
    \item \textbf{Protocol Coverage:} The current implementation focuses on HTTP-based attacks. Expansion to additional protocols (DNS, SMTP, FTP) would broaden the detection scope.
    \item \textbf{Longitudinal User Studies:} Longer-term evaluation of the educational component's influence on user behaviour would provide stronger evidence for sustained behavioural change.
    \item \textbf{Federated Learning:} Implementing privacy-preserving federated learning would enable threat intelligence sharing across organisations without exposing sensitive data.
    \item \textbf{Multilingual Support:} Extending the chatbot to support multiple languages would increase the platform's accessibility across different regions.
    \item \textbf{Adaptive Learning Paths:} Incorporating user proficiency tracking to personalise educational content delivery based on individual knowledge gaps.
\end{itemize}

% ---- 8. Conclusions ----
\section{Conclusions}
\label{sec:conclusions}

This paper has presented SentinelSphere, an innovative cybersecurity platform that successfully bridges the divide between advanced threat detection and human security awareness. Through integration with the ResilMesh framework and application of cutting-edge AI techniques, the platform delivers:

\begin{itemize}
    \item A 94\% F1 score in threat detection with a 69.5\% reduction in false positives, enabling SOC teams to concentrate on genuine threats.
    \item A 5.6$\times$ average performance improvement through Rust optimisation, with up to 326$\times$ speedup for batch processing.
    \item Successful processing of close to 11 million events in 30 minutes, demonstrating enterprise-scale capability.
    \item Empirically validated educational effectiveness, with 91.7\% chatbot engagement and 91.7\% Traffic Light comprehension among non-technical users.
\end{itemize}

SentinelSphere's dual approach, treating every security event as both a threat to mitigate and an opportunity to educate, represents a novel contribution to cybersecurity philosophy. By democratising security expertise through accessible AI tools, the platform addresses the critical skills gap while simultaneously strengthening organisational resilience against human-factor vulnerabilities.

Additionally, the successful integration with ResilMesh validates the importance of modular, extensible cybersecurity architectures that can accommodate innovative solutions while maintaining operational stability. This work extends the earlier contributions presented in~\cite{tantaroudas2026sentinelsphere_ares}, with the full peer-reviewed article available in~\cite{tantaroudas2026sentinelsphere_ore}. Future work will focus on expanding protocol coverage, implementing federated learning for privacy-preserving threat intelligence sharing, and conducting longitudinal studies to measure the sustained impact of integrated security education on organisational behaviour.

% ---- Acknowledgements ----
\section*{Acknowledgements}

This work was supported by the European Union's Horizon Europe programme under Grant Agreement No.\ 101119681. Views and opinions expressed are those of the authors only and do not necessarily reflect those of the European Union or the European Research Executive Agency. Neither the European Union nor the granting authority can be held responsible for them.

The authors acknowledge the contribution of the ResilMesh consortium partners for providing the foundational cybersecurity framework and messaging infrastructure. We thank the Pedagogy Department at the University of Western Macedonia for facilitating the educational workshop and all participants for their valuable feedback.

% ---- References ----


\begin{thebibliography}{36}

\bibitem[Ramezan(2023)]{ramezan2023}
Ramezan, C.A. (2023).
\newblock Examining the Cyber Skills Gap: An Analysis of Cybersecurity Positions by Sub-Field.
\newblock \textit{Journal of Information Systems Education}, 34(1):94--105.
\newblock \url{https://jise.org/Volume34/n1/JISE2023v34n1pp94-105.html}

\bibitem[{ISC2}(2023)]{isc2_2023}
{ISC2} (2023).
\newblock ISC2 Cybersecurity Workforce Study 2023.
\newblock International Information System Security Certification Consortium, Alexandria, VA.
\newblock \url{https://www.isc2.org/research}

\bibitem[Kucuk and Uysal(2022)]{kucuk2022}
Kucuk, M.F. and Uysal, I. (2022).
\newblock Anomaly Detection in Self-Organizing Networks: Conventional Versus Contemporary Machine Learning.
\newblock \textit{IEEE Access}, 10:61744--61752.
\newblock \doi{10.1109/ACCESS.2022.3181910}

\bibitem[{Ponemon Institute}(2023)]{ponemon2023}
{Ponemon Institute} (2023).
\newblock The Cost of Malware Containment.
\newblock Ponemon Institute LLC, Traverse City, MI.

\bibitem[{Verizon}(2025)]{verizon2025}
{Verizon} (2025).
\newblock 2025 Data Breach Investigations Report.
\newblock Verizon Business, New York.
\newblock \url{https://www.verizon.com/business/resources/reports/dbir/}

\bibitem[{ENISA}(2023)]{enisa2023}
{ENISA} (2023).
\newblock ENISA Threat Landscape 2023.
\newblock European Union Agency for Cybersecurity, Luxembourg.
\newblock \doi{10.2824/782573}

\bibitem[{NIST}(2024)]{nist2024}
{National Institute of Standards and Technology} (2024).
\newblock The NIST Cybersecurity Framework (CSF) 2.0.
\newblock NIST Cybersecurity White Paper, U.S. Department of Commerce, Washington DC.
\newblock \doi{10.6028/NIST.CSWP.29}

\bibitem[{ResilMesh Consortium}(2023)]{resilmesh2023}
{ResilMesh Consortium} (2023).
\newblock ResilMesh: Situation Aware enabled Cyber Resilience for Dispersed, Heterogenous Cyber Systems.
\newblock EU Horizon Europe Project 101119681, Technical Report.

\bibitem[Nguyen et~al.(2024)]{nguyen2024}
Nguyen, T., Sipola, T., and Hautam{\"a}ki, J. (2024).
\newblock Machine Learning Applications of Quantum Computing: A Review.
\newblock In \textit{Proceedings of the 23rd European Conference on Cyber Warfare and Security (ECCWS 2024)}, pp.\ 370--377. Academic Conferences International.
\newblock \doi{10.34190/eccws.23.1.2390}

\bibitem[Sadlek et~al.(2024)]{sadlek2024}
Sadlek, L., Hus{\'a}k, M., and {\v{C}}eleda, P. (2024).
\newblock Hierarchical Modeling of Cyber Assets in Kill Chain Attack Graphs.
\newblock In \textit{2024 20th International Conference on Network and Service Management (CNSM)}, pp.\ 1--7. IEEE.
\newblock \doi{10.23919/CNSM62983.2024}

\bibitem[Tantaroudas et~al.(2026a)]{tantaroudas2026sentinelsphere_ares}
Tantaroudas, N.D., Karachalios, I., and McCracken, A.J. (2026).
\newblock SentinelSphere: AI-Driven Cybersecurity Platform Combining Threat Detection with Security Awareness.
\newblock In \textit{Proceedings of the 21st International Conference on Availability, Reliability and Security (ARES 2025)}. ACM.
\newblock \doi{10.1145/3664476.3670446}

\bibitem[Sommer and Paxson(2010)]{sommer2010}
Sommer, R. and Paxson, V. (2010).
\newblock Outside the Closed World: On Using Machine Learning for Network Intrusion Detection.
\newblock In \textit{2010 IEEE Symposium on Security and Privacy}, pp.\ 305--316. IEEE.
\newblock \doi{10.1109/SP.2010.25}

\bibitem[Yin et~al.(2017)]{yin2017}
Yin, C., Zhu, Y., Fei, J., and He, X. (2017).
\newblock A Deep Learning Approach for Intrusion Detection Using Recurrent Neural Networks.
\newblock \textit{IEEE Access}, 5:21954--21961.
\newblock \doi{10.1109/ACCESS.2017.2762418}

\bibitem[Vinayakumar et~al.(2019)]{vinayakumar2019}
Vinayakumar, R., Alazab, M., Soman, K.P., Poornachandran, P., Al-Nemrat, A., and Venkatraman, S. (2019).
\newblock Deep Learning Approach for Intelligent Intrusion Detection System.
\newblock \textit{IEEE Access}, 7:41525--41550.
\newblock \doi{10.1109/ACCESS.2019.2895334}

\bibitem[Sharafaldin et~al.(2018a)]{sharafaldin2018a}
Sharafaldin, I., Lashkari, A.H., and Ghorbani, A.A. (2018).
\newblock Toward Generating a New Intrusion Detection Dataset and Intrusion Traffic Characterisation.
\newblock In \textit{Proceedings of the 4th International Conference on Information Systems Security and Privacy (ICISSP)}, pp.\ 108--116.
\newblock \doi{10.5220/0006639801080116}

\bibitem[Sharafaldin et~al.(2019)]{sharafaldin2019}
Sharafaldin, I., Lashkari, A.H., Hakak, S., and Ghorbani, A.A. (2019).
\newblock Developing Realistic Distributed Denial of Service (DDoS) Attack Dataset and Taxonomy.
\newblock In \textit{2019 International Carnahan Conference on Security Technology (ICCST)}, pp.\ 1--8. IEEE.
\newblock \doi{10.1109/CCST.2019.8888419}

\bibitem[Sharafaldin et~al.(2018b)]{sharafaldin2018b}
Sharafaldin, I., Lashkari, A.H., and Ghorbani, A.A. (2018).
\newblock A Detailed Analysis of the CICIDS2017 Data Set.
\newblock In \textit{Information Systems Security and Privacy}, pp.\ 172--188. Springer.
\newblock \doi{10.1007/978-3-030-25109-3_9}

\bibitem[Khan et~al.(2023)]{khan2023}
Khan, M.A., Iqbal, N., Jamil, H., and Kim, D.-H. (2023).
\newblock An optimised ensemble prediction model using AutoML based on soft voting classifier for network intrusion detection.
\newblock \textit{Journal of Network and Computer Applications}, 212:103560.
\newblock \doi{10.1016/j.jnca.2022.103560}

\bibitem[Ferrag et~al.(2020)]{ferrag2020}
Ferrag, M.A., Maglaras, L., Moschoyiannis, S., and Janicke, H. (2020).
\newblock Deep Learning for Cyber Security Intrusion Detection: Approaches, Datasets, and Comparative Study.
\newblock \textit{Journal of Information Security and Applications}, 50:102419.
\newblock \doi{10.1016/j.jisa.2019.102419}

\bibitem[Aktar and Nur(2023)]{aktar2023}
Aktar, S. and Nur, A.Y. (2023).
\newblock Towards DDoS attack detection using deep learning approach.
\newblock \textit{Computers \& Security}, 129:103251.
\newblock \doi{10.1016/j.cose.2023.103251}

\bibitem[Motlagh et~al.(2024)]{motlagh2024}
Motlagh, F.N., Hajizadeh, M., Majd, M., Najafi, P., Cheng, F., and Meinel, C. (2024).
\newblock Large Language Models in Cybersecurity: State-of-the-Art.
\newblock \textit{arXiv preprint arXiv:2402.00891}.
\newblock \doi{10.48550/arXiv.2402.00891}

\bibitem[Xu et~al.(2024)]{xu2024}
Xu, H., Wang, S., Li, N., Zhao, Y., Chen, K., Wang, K., Liu, Y., Yu, T., and Wang, H. (2024).
\newblock Large Language Models for Cyber Security: A Systematic Literature Review.
\newblock \textit{arXiv preprint arXiv:2405.04760}.
\newblock \doi{10.48550/arXiv.2405.04760}

\bibitem[Zhang et~al.(2025)]{zhang2025}
Zhang, J., Bu, H., Wen, H., Liu, Y., Fei, H., Xi, R., Li, L., Yang, Y., Zhu, H., and Meng, D. (2025).
\newblock When LLMs Meet Cybersecurity: A Systematic Literature Review.
\newblock \textit{Cybersecurity}, 8:55.
\newblock \doi{10.1186/s42400-025-00357-y}

\bibitem[Jaffal et~al.(2025)]{jaffal2025}
Jaffal, N.O., Alkhanafseh, M., and Mohaisen, D. (2025).
\newblock Large Language Models in Cybersecurity: A Survey of Applications, Vulnerabilities, and Defense Techniques.
\newblock \textit{AI}, 6(9):216.
\newblock \doi{10.3390/ai6090216}

\bibitem[Atlam(2025)]{atlam2025}
Atlam, H.F. (2025).
\newblock LLMs in Cyber Security: Bridging Practice and Education.
\newblock \textit{Big Data and Cognitive Computing}, 9(7):184.
\newblock \doi{10.3390/bdcc9070184}

\bibitem[Hassanin and Moustafa(2024)]{hassanin2024}
Hassanin, M. and Moustafa, N. (2024).
\newblock A Comprehensive Overview of Large Language Models (LLMs) for Cyber Defences: Opportunities and Directions.
\newblock \textit{arXiv preprint arXiv:2405.14487}.
\newblock \doi{10.48550/arXiv.2405.14487}

\bibitem[Chhetri(2024)]{chhetri2024}
Chhetri, C. (2024).
\newblock Exploring Large Language Model-Powered Pedagogical Approaches to Cybersecurity Education.
\newblock In \textit{Proceedings of the 25th Annual Conference on Information Technology Education (SIGITE '24)}, pp.\ 115--120. ACM.
\newblock \doi{10.1145/3686612.3686643}

\bibitem[Hadlington(2017)]{hadlington2017}
Hadlington, L. (2017).
\newblock Human Factors in Cybersecurity: Examining the Link Between Internet Addiction, Impulsivity, Attitudes Towards Cybersecurity, and Risky Cybersecurity Behaviours.
\newblock \textit{Heliyon}, 3(7):e00346.
\newblock \doi{10.1016/j.heliyon.2017.e00346}

\bibitem[Aldawood and Skinner(2019)]{aldawood2019}
Aldawood, H. and Skinner, G. (2019).
\newblock Reviewing Cyber Security Social Engineering Training and Awareness Programs---Pitfalls and Ongoing Issues.
\newblock \textit{Future Internet}, 11(3):73.
\newblock \doi{10.3390/fi11030073}

\bibitem[Bada et~al.(2019)]{bada2019}
Bada, M., Sasse, A.M., and Nurse, J.R. (2019).
\newblock Cyber Security Awareness Campaigns: Why do they fail to change behaviour?
\newblock \textit{arXiv preprint arXiv:1901.02672}.
\newblock \doi{10.48550/arXiv.1901.02672}

\bibitem[Zhang-Kennedy and Chiasson(2021)]{zhangkennedy2021}
Zhang-Kennedy, L. and Chiasson, S. (2021).
\newblock A Systematic Review of Multimedia Tools for Cybersecurity Awareness and Education.
\newblock \textit{ACM Computing Surveys}, 54(1):12.
\newblock \doi{10.1145/3427920}

\bibitem[Araujo et~al.(2024)]{araujo2024}
Araujo, M.S.d., Machado, B.A.S., and Passos, F.U. (2024).
\newblock Resilience in the Context of Cyber Security: A Review of the Fundamental Concepts and Relevance.
\newblock \textit{Applied Sciences}, 14(5):2116.
\newblock \doi{10.3390/app14052116}

\bibitem[Ramos-Cruz et~al.(2024)]{ramos2024}
Ramos-Cruz, B., Andreu-Perez, J., and Mart{\'\i}nez, L. (2024).
\newblock The Cybersecurity Mesh: A Comprehensive Survey of Involved Artificial Intelligence Methods, Cryptographic Protocols and Challenges for Future Research.
\newblock \textit{arXiv preprint arXiv:2402.18373}.
\newblock \doi{10.48550/arXiv.2402.18373}

\bibitem[Somma et~al.(2024)]{somma2024}
Somma, M., Flatscher, A., and Stojanovi{\'c}, B. (2024).
\newblock Edge-Based Anomaly Detection: Enhancing Performance and Sustainability in Smart Water Distribution Systems.
\newblock In \textit{2024 32nd Telecommunications Forum (TELFOR)}, pp.\ 1--4. IEEE.
\newblock \doi{10.1109/TELFOR63250.2024}

\bibitem[Bhatt et~al.(2014)]{bhatt2014}
Bhatt, S., Manadhata, P.K., and Zomlot, L. (2014).
\newblock The Operational Role of Security Information and Event Management Systems.
\newblock \textit{IEEE Security \& Privacy}, 12(5):35--41.
\newblock \doi{10.1109/MSP.2014.103}

\bibitem[Tantaroudas et~al.(2026b)]{tantaroudas2026sentinelsphere_ore}
Tantaroudas, N.D., Karachalios, I., and McCracken, A.J. (2026).
\newblock SentinelSphere: An AI-driven cybersecurity platform integrating real-time threat detection with security awareness education [version 1; peer review: 2 approved with reservations].
\newblock \textit{Open Research Europe}, 6:58.
\newblock \doi{10.12688/openreseurope.22957.1}

\end{thebibliography}
\end{document}